\documentclass[graybox]{SpringerStyles/svmult}

\usepackage{mathptmx}       
\usepackage{helvet}         
\usepackage{courier}        
\usepackage{type1cm}        
\usepackage{makeidx}         
\usepackage{graphicx}        
\usepackage{multicol}        
\usepackage[bottom]{footmisc}

\makeindex             

\usepackage{amsmath,amsfonts}
\usepackage{listings}
\usepackage{url}
\usepackage{subfigure}
\newcommand{\reffig}[1]{{Fig.~\ref{#1}}}

\newcommand{\refsec}[1]{{Sect.~\ref{#1}}}
\newcommand{\reftab}[1]{{Table~\ref{#1}}}

\newcommand{\refeqn}[1]{{(\ref{#1})}}

\begin{document}

{
\title*{Efficient Expression Templates for Operator Overloading-based Automatic Differentiation}
\titlerunning{Efficient Expression Templates}
\author{Eric Phipps \and Roger Pawlowski}
\institute{
Eric Phipps \at Sandia National Laboratories\textsuperscript{\dag}, Optimization and Uncertainty Quantification Department, Albuquerque, NM, USA, \url{etphipp@sandia.gov}
\and 
Roger Pawlowski \at Sandia National Laboratories\textsuperscript{\dag}, Multiphysics Simulation Technologies Department, Albuquerque, NM, USA \url{rppawlo@sandia.gov}
\and
\textsuperscript{\dag}Sandia National Laboratories is a multi-program laboratory managed and operated by Sandia Corporation, a wholly owned subsidiary of Lockheed Martin Corporation, for the U.S. Department of Energy's National Nuclear Security Administration under contract DE-AC04-94AL85000.}
\maketitle

\abstract{Expression templates are a well-known set of techniques for improving the efficiency of operator overloading-based forward mode automatic differentiation schemes in the C++ programming language by translating the differentiation from individual operators to whole expressions.  However standard expression template approaches result in a large amount of duplicate computation, particularly for large expression trees, degrading their performance.  In this paper we describe several techniques for improving the efficiency of expression templates and their implementation in the automatic differentiation package Sacado~\cite{Phipps2008LST,PaperXXSacadoURL}.  We demonstrate their improved efficiency through test functions as well as their application to differentiation of a large-scale fluid dynamics simulation code.
\keywords{Forward mode, operator overloading, expression templates,  C++}
}

\lstset{language=C++, basicstyle=\small, keywordstyle=\textrm}

\section{Introduction}\label{PaperXXsec:intro}

Automatic differentiation (AD) techniques for compiled languages such as C++ and Fortran fall generally into two basic categories:  source transformation and operator overloading.  Source transformation involves a preprocessor that reads and parses the code to be differentiated, applies differentiation rules to this code, and generates new source code for the resulting derivative calculation that can be compiled along with the rest of the undifferentiated source code.  This approach is quite popular for simpler languages such as Fortran and C, however is challenging for C++ due to the complexity of the language.  An alternative approach for C++ (and many other languages) is operator overloading.  Here new derived types storing derivative values and corresponding overloaded operators are created so that when the fundamental scalar type in the calculation (\lstinline{float} or \lstinline{double}) is replaced by these new types and evaluated, the relevant derivatives are computed as a side-effect.  This approach is attractive in that it uses native features of the language, making operator overloading-based AD tools simple to implement and use.  There are two basic challenges for operator overloading schemes however:  run-time efficiency and facilitating the necessary type change from the floating point type to AD types.  For forward mode AD, expression templates can be used to partially address the first of these challenges.  However achieving the full performance benefits of expression templates is challenging and is the subject of this paper.  For the second challenge, we advocate a templating-based approach which has been described elsewhere\cite{Bartlett2006ADo,PaperXXTBGP1,PaperXXTBGP2,Phipps2008LST}.  

This paper is organized as follows.  We first describe standard expression template techniques and their application to forward mode automatic differentiation in ~\refsec{PaperXXsec:et}.  For concreteness, we describe the simple implementation of these techniques in the AD package Sacado~\cite{Phipps2008LST,PaperXXSacadoURL}.  Then in~\refsec{PaperXXsec:improved} we describe two techniques for improving the performance of expression templates:  caching and expression-level reverse mode.  We demonstrate significantly improved performance for these techniques, particularly for large expressions, by applying them to two test functions.  We then briefly describe applying all of these techniques to a large-scale fluid dynamics simulation in~\refsec{PaperXXsec:application}, again demonstrating improved performance on a real-world application.  We then close with several concluding remarks in~\refsec{PaperXXsec:conclusions}.

\section{Expression Templates for Forward Mode AD}\label{PaperXXsec:et}

As described above, operator overloading-based AD schemes work by first creating a new derived type and corresponding overloaded operators.  While operator overloading can be used for any AD mode, and there are many ways of implementing the overloaded operators for any given AD mode, we will restrict this discussion to tapeless implementations of the first-order vector forward mode.  Here the AD type typically contains a floating point value to represent the value of an intermediate variable, and an array to store the derivatives of that intermediate variable with respect to the independent variables (see~\cite{Griewank2000EDP} for an introduction to basic AD implementations).  The implementation of each overloaded operator then involves calculation of the value of that operation from the values of the arguments and stored in the value of the result, and a loop over the derivative components using the corresponding derivative rule from basic differential calculus.  

There are two basic problems with this approach.  First, each intermediate operation within an expression requires creation of at least one temporary object, and creating and destroying this object adds significant run-time overhead.  Second, the AD implementation is limited to differentiating one operation at a time, each involving a loop over derivative components.  Together these problems often result in inefficient derivative code.  Expression templates~\cite{PaperXXVelhuizenET} are a technique that can address these issues.  They were first used for AD in the Fad package~\cite{Aubert2002ETa,Aubert2001ADi}, and later incorporated into Sacado.  Here the AD type is fundamentally the same, however the operators return an object encoding the operation type and a handle to their arguments, instead of directly evaluating the derivative.  As each term in the expression is evaluated, a tree is created encoding the structure of the whole expression.  Then the assignment operator for the AD type loops through this tree recursively applying the chain rule.  An implementation of these ideas for the $\times$ operator is shown below.   
\begin{lstlisting}[caption=Partial expression template-based operator overloading implementation.,label=PaperXXfig:simple_et,basicstyle=\scriptsize]
// Expression template-based Forward AD type
template <typename T> class Expr {};
class ETFadTag {};
class ETFad : public Expr<ETFadTag> {
  double val_;              // value
  std::vector<double> dx_;  // derivatives
public:  
  explicit ETFad(int N) : val_(0), dx_(N) {} // Constructor
  int size() const { return dx_.size(); } 
  double  val() const { return val_; } // Return value
  double& val()       { return val_; } // Return value
  double  dx(int i) const { return dx_[i]; } // Return derivative
  double& dx(int i)       { return dx_[i]; } // Return derivative
  
  // Expression template assignment operator
  template <typename T> ETFad& operator=(const Expr<T>& x)  {
    val_ = x.val();
    dx_.resize(x_.size());
    for (int i=0; i<x.size(); i++)
      dx_[i] = x.dx(i);
  }
};

// Specialization of Expr to multiplication
template <typename ExprT1, typename ExprT2>  class MultTag {};
template <typename T1, typename T2> class Expr<MultTag<Expr<T1>, Expr<T2> > > {
  const Expr<T1>& a_; const Expr<T2>& b_;
public:
  Expr(const Expr<T1>& a, const Expr<T2>& b) : a_(a), b_(b) {}
  int size() const { return a.size(); }
  double val() const { return a.val() * b.val(); }
  double dx(int i) const { return a.val()*b.dx(i)+a.dx(i)*b.val(); }
};
  
// Expression template implementation of a*b
template <typename T1, typename T2> Expr< MultTag< Expr<T1>, Expr<T2> > > 
operator * (const Expr<T1>& a, const Expr<T2>& b) {
  return Expr< MultTag< Expr<T1>, Expr<T2> > >(a,b);
}
\end{lstlisting}
Each overloaded operator (\lstinline{operator*()} in this case) returns a simple expression object that stores just references to its arguments with the kind of operation encoded in the type of the expression (though \lstinline{MultTag} in this case).  Notice that the overloaded multiplication operator takes general expressions as arguments, and thus an {\em expression tree} is generated by each term in any given expression.  The class \lstinline{ETFad} stores the value and derivatives of any given intermediate variable, and also being derived from a specialization of the \lstinline{Expr}, is a leaf node in an expression tree.  The tree is then traversed to actually compute the value of the expression and its derivatives through the template assignment operator (\lstinline{ETFad::operator=()}) when the expression is assigned to an \lstinline{ETFad} intermediate variable.  Since the nodes in the expression tree (such as \lstinline{Expr< MultTag< > >}) just contain references,  a good optimizing compiler can often eliminate them all together and generate code that is functionally equivalent to that shown below when applied to $d=a\times b\times c$.  
\begin{lstlisting}[caption=Equivalent derivative code resulting from differentiation of $a\times b\times c$.,label=PaperXXfig:equiv_deriv,basicstyle=\scriptsize]
d.val() = a.val()*b.val()*c.val();
for (int i=0; i<d.size(); i++)
  d.dx(i) = (a.val()*b.val())*c.dx(i)+(a.val()*b.dx(i)+a.dx(i)*b.val())*c.val();
\end{lstlisting}
Thus all of the intermediate temporary AD objects have been removed and the loops have been fused into a single loop over the derivative components for $d$.  This often removes much of the overhead associated with a simple operator overloading approach.  We note that constants and passive variables introduce additional complexity into the implementation which is not shown or discussed here.  Also, a dynamically allocated derivative array was chosen to allow the number of derivative components to be determined at compile time.  In cases when this is known at compile time, a fixed-length array can be chosen to improve performance.

\section{Improving Performance of Expression Templates}\label{PaperXXsec:improved}

While the expression template approach can significantly reduce the overhead associated with operator overloading, there is still room for improvement in reducing the cost of the differentiation.  Careful examination of Listing~\ref{PaperXXfig:equiv_deriv} reveals a basic problem:  the calculation of the value portion of intermediate terms in the expression can be repeated multiple times.  This is particularly troublesome for large expressions involving many terms or expressions involving transcendental functions whose values are expensive to compute.  In theory, a good optimizing compiler should be able to remove these redundant calculations through common sub-expression elimination, however our experience has been that no compiler actually does (which is supported by the numerical experiments below).

To remedy this, the compiler must be coerced into computing any needed values just once for all of the intermediate operations in the expression tree.  We have investigated overcoming this problem by caching the value and/or partial derivatives of each intermediate operation in the expression objects themselves.  The small modifications to the multiplication expression template from Listing~\ref{PaperXXfig:simple_et} are shown below where the \lstinline{cache()} method in this case stores the values of the operator arguments \lstinline{a} and \lstinline{b} (which in this special case is both the values and partial derivatives of the arguments).  These cached values are then used in any subsequent calls to \lstinline{val()} or \lstinline{dx()}.  Since an expression class may be copied many times during the creation of a full expression-template, the caching computation is deferred until the expression template construction is complete by modifying the top-level expression-template assignment operator to call \lstinline{cache()} before any calls to \lstinline{val()} or \lstinline{dx()} (as opposed to caching within the expression constructors).  This approach eliminates the duplicate computation of intermediate values, at the expense of more complicated expression objects that the compiler may not be able to optimize away (including the now non-trivial copy constructors).  Nonetheless we have found this approach more efficient for recent compilers that support aggressive C++ optimization.
\begin{lstlisting}[caption=Modifications from Listing~\ref{PaperXXfig:simple_et} for caching expression template-based operator overloading.  Only the modifications from Listing~\ref{PaperXXfig:simple_et} are presented.  For brevity the simple modification to the ETFad assignment operator is suppressed.,label=PaperXXfig:cache_et,basicstyle=\scriptsize]
template <typename T1, typename T2> class Expr<MultTag<Expr<T1>, Expr<T2> > > {
  const Expr<T1>& a_; const Expr<T2>& b_;
  mutable double a_val_, b_val_;
public:
  Expr(const Expr<T1>& a, const Expr<T2>& b) : a_(a), b_(b) {}
  void cache() const { 
    a_.cache(); b_.cache(); a_val_ = a_.val(); b_val_ = b_.val(); 
  }
  double val() const { return a_val_*b_val_; }
  double dx(int i) const { return a_val_*b_.dx(i)+a_.dx(i)*b_val_; }
};
\end{lstlisting}

A second technique that can be used to generally improve the performance of forward mode AD is expression-level reverse mode~\cite{Bischof1996HAt}.  This results from the recognition that while derivatives are generally being propagated forward through the calculation, any given expression likely has multiple inputs and only one output.  Thus it should be more efficient to compute the derivative of the expression outputs with respect to its inputs using reverse mode AD and then combine those derivatives with the derivatives of the inputs using the chain rule.  This technique is common in source transformation tools such as ADIFOR~\cite{Bischof1996AAD}, and the ADTAGEO~\cite{Riehme2009ADT} tool implemented an instant graph elimination technique in an operator overloading fashion that is equivalent to expression-level reverse mode.  However we are unaware of any use of this approach in expression template-based approaches.  The challenge is implementing the technique in a way that can be effectively optimized by the compiler.

We have implemented expression-level reverse mode within the forward mode classes in our tool Sacado using template meta-programming techniques~\cite{PaperXXBoostMPLBook2004} similar to those found in the Boost MPL library~\cite{PaperXXBoostLib}.  Referring to Listing~\ref{PaperXXfig:elr_et_1}, the total number of arguments to the expression is accumulated in the \lstinline{num_args} member\footnote{Note that \lstinline{num_args} is a compile-time constant that is uniquely determined by each expression in the code, and thus while it is ``static'', it isn't a static variable in the traditional sense.} of each expression class as the expression tree is built up.  Leaves in the tree (objects of type \lstinline{ELRFad}) are treated as single argument identity functions.  Each binary operation such as \lstinline{operator*()} from Listing~\ref{PaperXXfig:elr_et_1} treats its full set of expression arguments as the union of its two arguments, thus the operation $a\times a$ would be treated as having two arguments.  The \lstinline{computePartials()} method for each expression class computes the partial derivatives of the result of that operation with respect to the expression arguments using reverse mode AD (\lstinline{bar} stores the derivative of the expression result with respect to that intermediate variable/operation).  These are stored in the \lstinline{partials} array, whose length is determined by \lstinline{num_args}, with the partials arising from the first argument in a binary operation stored in the first \lstinline{num_args1} locations and those from the second argument in the remaining \lstinline{num_args2} locations.  Then the arguments of the expression tree are returned by the \lstinline{getArg()} method allowing extraction of their derivative components.  
\begin{lstlisting}[caption=Additional expression template interface incorporating expression-level reverse mode.,label=PaperXXfig:elr_et_1,basicstyle=\scriptsize]
class ETFad : public Expr<ETFadTag> {
public: 
  static const int num_args = 1; // Number of expression args
  
  // Return partials w.r.t. arguments
  void computePartials(double bar, double partials[]) const { 
    partials[0] = bar;  }
    
  // Return argument Arg of expression
  template <typename Arg> const ETFad& getArg() const { return *this; }
};

template <typename T1, typename T2> class Expr<MultTag<Expr<T1>, Expr<T2> > > {
public:
  // Number of arguments to expression
  static const int num_args1 = Expr<T1>::num_args;
  static const int num_args2 = Expr<T2>::num_args;
  static const int num_args = num_args1 + num_args2;
  
  // Compute partial derivatives w.r.t. arguments
  void computePartials(double bar, double partials[]) const {
    a_.computePartials(bar*b_.val(), partials);
    b_.computePartials(bar*a_.val(), partials+num_args1);
  }
  
  // Return argument Arg for expression
  template <int Arg> const ETFad& getArg() const {
    if (Arg < num_args1) return a_.template getArg<Arg>();
    else return b_.template getArg<Arg-num_args1>();
  }	
};

\end{lstlisting}

These methods are then used to combine the expression-level reverse mode with the overall forward AD propagation through the new implementation of the assignment operator shown in Listing~\ref{PaperXXfig:elr_et_2}.  First the derivatives with respect to the expression arguments are computed using reverse mode AD.  These are then combined with the derivative components of the expression arguments using the functor \lstinline{LocalAccumOp} and the MPL function \lstinline{for_each}.  The overloaded \lstinline{operator()} of \lstinline{LocalAccumOp} computes the contribution of expression argument \lstinline{Arg} to final derivative component \lstinline{i} using the chain rule.  The MPL function \lstinline{for_each} then iterates over all of the expression arguments by iterating through the integral range $[0,M)$ where $M$ is the number of expression arguments.  Since $M$ is a compile-time constant and \lstinline{for_each} uses template recursion to perform the iteration, this effectively an unrolled loop.
\begin{lstlisting}[caption=Expression template forward AD propagation using expression-level reverse mode.,label=PaperXXfig:elr_et_2,basicstyle=\scriptsize]
// Functor for mpl::for_each to multiply partials and tangents
template <typename ExprT> struct LocalAccumOp {
  const ExprT& x_;
  mutable double t_;
  double partials_[Expr<T>::num_args];
  int i;
  LocalAccumOp(const ExprT& x_) : x(x_) {}
  template <typename ArgT> void operator () (ArgT arg) const {
    const int Arg = ArgT::value;
    t += partials[Arg] * x.template getArg<Arg>().dx(i);
  }
};

class ETFad : public Expr<ETFadTag> {
public: 
  // ELR expression template assignment operator
  template <typename T> ELRFad& operator=(const Expr<T>& x)  {
    val_ = x.val();
    dx_.resize(x.size());

    // Compute partials w.r.t. expression arguments
    LocalAccumOp< Expr<T> > op(x);
    x.computePartials(1.0, op.partials); 
    
    // Multiply partials with derivatives of arguments
    const int M = Expr<T>::num_args;
    for(op.i=0; op.i<x.size(); ++op.i) {
      op.t = 0.0;
      mpl::for_each< mpl::range_c< int, 0, M > > f(op);
      dx_[i] = op.t;
    }
    return *this;
  }
};
\end{lstlisting}

Note that as in the simple expression template implementation above, the value of each intermediate operation in the expression tree may be computed multiple times.  However the values are only computed in the \lstinline{computePartials()} and \lstinline{val()} methods, which are each only called once per expression, and thus the amount of re-computation only depends on the expression size, not the number of derivative components.  Clearly the caching approach discussed above can also be incorporated with the expression-level reverse mode approach, which will not be shown here.  

To test the performance of the various approaches, we apply them to 
\begin{center}
\vspace{-0.18in}
\begin{minipage}{3cm}
\begin{equation}
  y = \prod_{i=1}^M x_i \label{PaperXXeq:mult}
\end{equation}
\end{minipage}
\hspace{0.25cm} and
\begin{minipage}[b]{5cm}
\begin{equation}
  y = \overbrace{\sin(\sin(\dots\sin}^{\text{$M$ times}} (x)\dots)) \label{PaperXXeq:nested}
\end{equation}
\end{minipage}
\end{center}
for $M = 1, 2, 3, 4, 5, 10, 15, 20$.  Test function~\refeqn{PaperXXeq:mult} tests wide but shallow expressions, whereas function~\refeqn{PaperXXeq:nested} tests deep but narrow expressions, and together they are the extremes for expressions seen in any given computation.  In \reffig{PaperXXfig:times} we show the scaled run time (average wall clock time divided by the average undifferentiated expression evaluation time times the number of derivative components $N$)  of propagating $N=5$ and $N=50$ derivative components through these expressions for each value of $M$ using the standard expression template, expression-level reverse mode, caching, and caching expression-level reverse mode approaches implemented in Sacado.  Also included in these plots is a simple (tapeless) forward AD implementation without expression templates.  These tests were conducted using Intel 12.0 and GNU 4.5.3 compilers using standard aggressive optimization options (-O3), run on a single core of an Intel quad-core processor.  The GNU and Intel results were qualitatively similar with the GNU results shown here.  One can see that for a larger number of derivative components or expressions with transcendental terms, the standard expression-template approach performs quite poorly due to the large amount of re-computation.  All three of caching, expression-level reverse mode, and caching expression-level reverse mode are significant improvements, with the latter generally being the most efficient.   Moreover, even for expressions with one transcendental term but many derivative components, these approaches are a significant improvement.  For small expression sizes with no transcendental terms and few derivative components however, the differences are not significant.  Since most applications would likely consist primarily of small expressions with a mixture of algebraic and transcendental terms, we would still expect to see some improvement.
\begin{figure}[htb]
\begin{center}
\subfigure[Multiply function~\refeqn{PaperXXeq:mult} for $N=5$.]{\includegraphics[scale=1.0]{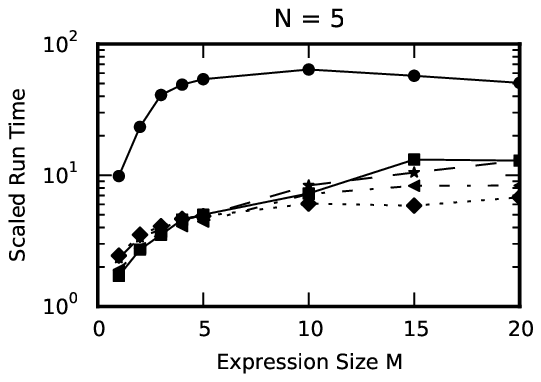}\label{PaperXXfig:mult5}}
\subfigure[Multiply function~\refeqn{PaperXXeq:mult} for $N=50$.]{\includegraphics[scale=1.0]{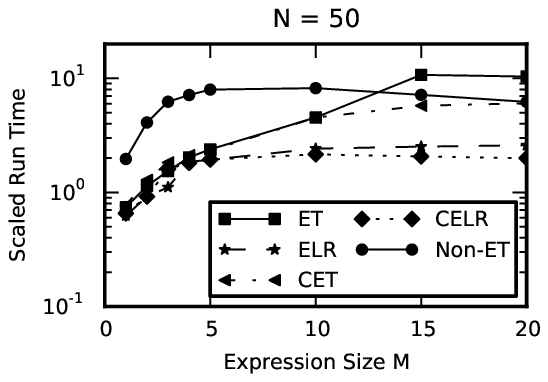}\label{PaperXXfig:mult50}}
\subfigure[Nested function~\refeqn{PaperXXeq:nested} for $N=5$.]{\includegraphics[scale=1.0]{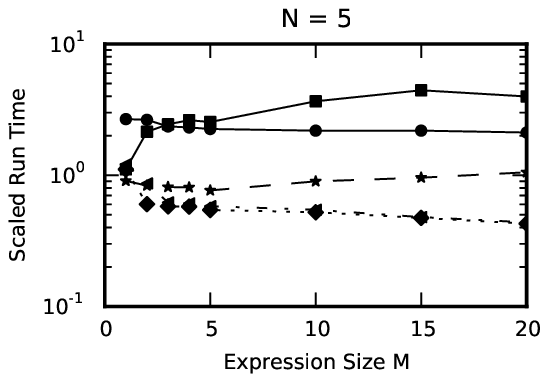}\label{PaperXXfig:nested5}}
\subfigure[Nested function~\refeqn{PaperXXeq:nested} for $N=50$.]{\includegraphics[scale=1.0]{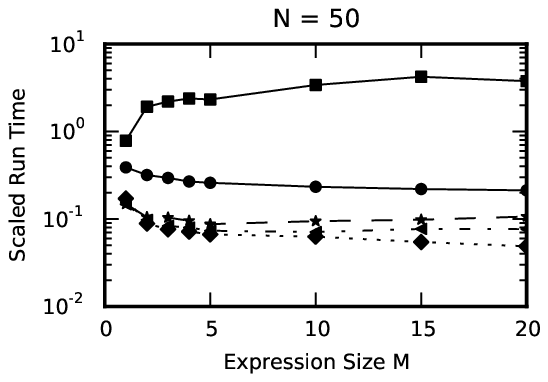}\label{PaperXXfig:nested50}}
\end{center}
\caption{Scaled derivative propagation time for expressions of various sizes.  Here ET refers to standard expression templates, ELR to expression-level reverse mode, CET/CELR to caching versions of these approaches, and Non-ET to an implementation without expression templates.}\label{PaperXXfig:times}
\end{figure}

\section{Application to Differentiation of a Fluid Dynamics Simulation}\label{PaperXXsec:application}

To demonstrate the impact of these approaches to problems of practical interest, we apply them to the problem of computing a steady-state solution to the decomposition of dilute species in a duct flow.  The problem is modeled by a system of coupled differential algebraic equations that enforce the conservation of momentum, energy, and mass under non-equilibrium chemical reaction.  The complete set of equations, the discretization technique and the solution algorithms are described in detail in \cite{PaperXXShadidCMAME2006}.  The system is discretized using a stabilized Galerkin finite element approach on an unstructured hexahedral mesh of 8000 cells with a linear Lagrange basis.  We solve three momentum equations, a total continuity equation, an energy equation and five species conservation equations resulting in 10 total equations.  Due to the strongly coupled nonlinear nature of the problem, a fully coupled, implicit, globalized inexact Newton-based solve \cite{PaperXXWalker1994} is applied.  This requires the evaluation of the Jacobian sensitivity matrix for the nonlinear system.  An element-based automatic differentiation approach~\cite{Bartlett2006ADo,Phipps2008LST} is applied via template-based generic programming~\cite{PaperXXTBGP1,PaperXXTBGP2} and Sacado, resulting in 80 derivative components in each element computation.  The five species decomposition mechanism uses the Arrhenius equation for the temperature dependent kinetic rate, thus introducing transcendental functions via the the source terms for the species conservation equations.

\reftab{PaperXXtab:rxn_trans} shows the evaluation times for the global Jacobian required for each Newton step, scaled by the product of the residual evaluation time and the number of derivative components per element.  The calculation was run on 16 processor cores using MPI parallelism and version 4.5.3 of the GNU compilers and -O3 optimization flags.  As would be expected, both caching and expression-level reverse mode approaches are significant improvements.  
\begin{table}
\caption{Scaled Jacobian evaluation time for reaction/transport problem.}
\label{PaperXXtab:rxn_trans}
\begin{tabular}{lc}
\hline\noalign{\smallskip}
Implementation & Scaled Jacobian Evaluation Time \\
\noalign{\smallskip}\svhline\noalign{\smallskip}
Standard expression template     & 0.187 \\
Expression-level reverse         & 0.121 \\
Caching expression template      & 0.129 \\
Caching expression-level reverse & 0.120\\
\noalign{\smallskip}\hline\noalign{\smallskip}
\end{tabular}
\end{table}

\section{Concluding Remarks}\label{PaperXXsec:conclusions}

In this paper we described challenges for using expression template techniques in operator overloading-based implementations of forward mode AD in the C++ programming language, and two approaches for overcoming them:  caching and expression-level reverse mode.  While expression-level reverse mode is not a new idea, we believe our use of it in expression template approaches, and its implementation using template meta-programming is unique.  Together, these techniques significantly improve the performance of expression template approaches on a wide range of expressions, demonstrated through small test problems and application to a reacting flow fluid dynamics simulation.  In the future we are interested in applying the approach in~\cite{Naumann2008OVE} for accumulating the expression gradient even more efficiently, which should be feasible with general meta-programming techniques.  
\nocite{Corliss2002ADo,bischof2008aia}

\bibliographystyle{SpringerStyles/spmpsci}
\bibliography{ad,Papers/PaperXX/paper}
}

\end{document}